\documentclass[aps,prc,twocolumn,superscriptaddress,showpacs,showkeys]{revtex4}

\usepackage{graphicx}
\usepackage{dcolumn}
\usepackage{bm}

\usepackage{float}

\usepackage[usenames]{color}

\begin{document}


\title{Sequential and simultaneous emission of fragments from p+Al collisions}



\author{M.Fidelus}
\affiliation{M. Smoluchowski Institute of Physics, Jagiellonian
University, Reymonta 4, 30059 Krak{\'o}w, Poland}
\author{D.Filges}
\affiliation{J{\"u}lich Center for Hadron Physics, Forschungszentrum
J{\"u}lich, 52425 J{\"u}lich, Germany} \affiliation{Institut f{\"ur}
Kernphysik, Forschungszentrum J{\"u}lich, 52425 J{\"u}lich, Germany}
\author{F.Goldenbaum}
\affiliation{J{\"u}lich Center for Hadron Physics, Forschungszentrum
J{\"u}lich, 52425 J{\"u}lich, Germany} \affiliation{Institut f{\"ur}
Kernphysik, Forschungszentrum J{\"u}lich, 52425 J{\"u}lich, Germany}
\author{H.Hodde}
\affiliation{Institut f{\"ur} Strahlen- und Kernphysik, Bonn
University,  53121 Bonn, Germany}
\author{A.Jany}
\affiliation{M. Smoluchowski Institute of Physics, Jagiellonian
University, Reymonta 4, 30059 Krak{\'o}w, Poland}
\author{L.Jarczyk}
\affiliation{M. Smoluchowski Institute of Physics, Jagiellonian
University, Reymonta 4, 30059 Krak{\'o}w, Poland}
\author{B.Kamys}  \email[Corresponding author: ]{ufkamys@cyf-kr.edu.pl}
\affiliation{M. Smoluchowski Institute of Physics, Jagiellonian
University, Reymonta 4, 30059 Krak{\'o}w, Poland}
\author{M.Kistryn}
\affiliation{H. Niewodnicza{\'n}ski Institute of Nuclear Physics
PAN, Radzikowskiego 152, 31342 Krak{\'o}w, Poland}
\author{St.Kistryn}
\affiliation{M. Smoluchowski Institute of Physics, Jagiellonian
University, Reymonta 4, 30059 Krak{\'o}w, Poland}
\author{St.Kliczewski}
\affiliation{H. Niewodnicza{\'n}ski Institute of Nuclear Physics
PAN, Radzikowskiego 152, 31342 Krak{\'o}w, Poland}
\author{E.Kozik}
\affiliation{H. Niewodnicza{\'n}ski Institute of Nuclear Physics
PAN, Radzikowskiego 152, 31342 Krak{\'o}w, Poland}
\author{P.Kulessa}
\affiliation{H. Niewodnicza{\'n}ski Institute of Nuclear Physics
PAN, Radzikowskiego 152, 31342 Krak{\'o}w, Poland}
\author{H.Machner}
\affiliation{Universit{\"a}t Duisburg-Essen, Fakult{\"a}t f{\"u}r
Physik,  Lotharstr.1, 47048 Duisburg, Germany}
\author{A.Magiera}
\affiliation{M. Smoluchowski Institute of Physics, Jagiellonian
University, Reymonta 4, 30059 Krak{\'o}w, Poland}
\author{B.Piskor-Ignatowicz}
\affiliation{M. Smoluchowski Institute of Physics, Jagiellonian
University, Reymonta 4, 30059 Krak{\'o}w, Poland}
\affiliation{J{\"u}lich Center for Hadron Physics, Forschungszentrum
J{\"u}lich, 52425 J{\"u}lich, Germany} \affiliation{Institut f{\"ur}
Kernphysik, Forschungszentrum J{\"u}lich, 52425 J{\"u}lich, Germany}
\author{K.Pysz}
\affiliation{H. Niewodnicza{\'n}ski Institute of Nuclear Physics
PAN, Radzikowskiego 152, 31342 Krak{\'o}w, Poland}
\author{Z.Rudy}
\affiliation{M. Smoluchowski Institute of Physics, Jagiellonian
University, Reymonta 4, 30059 Krak{\'o}w, Poland}
\author{Sushil K. Sharma}
\affiliation{M. Smoluchowski Institute of Physics, Jagiellonian
University, Reymonta 4, 30059 Krak{\'o}w, Poland}
\affiliation{J{\"u}lich Center for Hadron Physics, Forschungszentrum
J{\"u}lich, 52425 J{\"u}lich, Germany} \affiliation{Institut f{\"ur}
Kernphysik, Forschungszentrum J{\"u}lich, 52425 J{\"u}lich, Germany}
\author{R.Siudak}
\affiliation{H. Niewodnicza{\'n}ski Institute of Nuclear Physics
PAN, Radzikowskiego 152, 31342 Krak{\'o}w, Poland}
\author{M.Wojciechowski}
\affiliation{M. Smoluchowski Institute of Physics, Jagiellonian
University, Reymonta 4, 30059 Krak{\'o}w, Poland}

\collaboration{PISA - \textbf{P}roton \textbf{I}nduced
\textbf{S}p\textbf{A}llation collaboration}

\date{\today}

\begin{abstract}
The energy and angular dependence of double differential cross
sections $d^{2}\sigma/d\Omega dE$ were measured for
$p,d,t,^{3,4,6}$He, $^{6,7,8}$Li, $^{7,9,10}$Be, and $^{10,11,12}$B
 produced in collisions of  1.2, 1.9, and 2.5 GeV protons with an Al  target. The
spectra and angular distributions of Li, Be, and B isotopes indicate
a presence of two contributions: an isotropic, low energy one which
is attributed to the evaporation of particles from excited remnants
of the intranuclear cascade, and an anisotropic part which is
interpreted to be due to multifragmentation of the remnants. It was
found that such a hypothesis leads to a very good description of
spectra and angular distributions  of all intermediate mass
fragments ($^6$He - $^{12}$B) using the critical value of the
excitation energy per nucleon as a single parameter, varying slowly
with the beam energy.

\end{abstract}



\maketitle


\section{\label{sec:introduction} Introduction}

In recent studies of  reactions induced by GeV protons on Au
\cite{BUB07A,BUD08A} and Ni targets \cite{BUD09A,BUD10A}, it has
been found that the inclusive spectra of intermediate mass fragments
(IMF),\emph{ i.e.}, particles with Z$\geq$ 3, but lighter than
fission fragments, as \emph{e.g.} Li, Be, B, etc., contain two
components which differ in energy and angular dependencies. The low
energy component of the spectra is almost angle independent, while
the high energy part of the spectra changes with the angle becoming
more steep with increasing scattering angle. These properties of the
spectra could not be quantitatively reproduced by the traditional
two step model which assumes that the first stage of the reaction
proceeds via an intranuclear cascade of the nucleon-nucleon and
nucleon-pion collisions leaving the excited target remnant in
equilibrium whereas the second stage consists in the evaporation of
nucleons and composite particles.

The same effect of the presence of two components in the spectra has
been observed for various target nuclei by other authors,\emph{
e.g.}, by Green \emph{et al.} \cite{GRE80A,GRE87A} for reactions  in
p+Ag system at $T_p$=0.21, 0.3, and 0.48 GeV, by Herbach\emph{ et
al.} \cite{HER06A} for target nuclei between Al and Th at 1.2 GeV,
by Letourneau \emph{et al.} \cite{LET02A} for p+Au collisions at 2.5
GeV, by Westfall et al. \cite{WES78A} for C, Al, Ag and U targets
irradiated by protons of 2.1 and 4.9 GeV energies, by Hyde \emph{et
al.} \cite{HYD71A} for p+Ag system at 5.5 GeV, and by Poskanzer
\emph{et al.} \cite{POS71A} for p+U reactions at 5.5 GeV.

The analysis of the energy and angular dependencies of differential
cross sections for IMFs from p+Au and p+Ni reactions
\cite{BUB07A,BUD08A,BUD09A,BUD10A} has shown that the data can be
well reproduced by a phenomenological model which assumes that the
composite particles are emitted isotropically from two sources
moving along the beam direction. 
%
%
%
The break-up of bombarded nuclei, occuring in the first stage of the
collision, was proposed \cite{BUB07A} to account for the observed
phenomenon.

The competitive hypothesis, which is studied in the present work,
states that the excited nuclei are divided into two groups not due
to the geometrical conditions imposed by their creation mode but on
the contrary, by their decay.  It may be conjectured that the decay
mode of the excited nuclei depends mainly on their excitation energy
per nucleon.  The nuclei which have small excitation energy
evaporate nucleons and composite particles, whereas the highly
excited nuclei (above some critical energy per nucleon) may undergo
a phase transition, i.e. multifragmentation appears
\cite{RIC01A,BOR08A}. The energy available in the multifragmentation
is usually larger than that in the evaporation process because the
multifragmentation appears only for highly excited nuclei.
Furthermore, the full excitation energy is then released in a single
act, whereas the evaporation consists of several stpdf in which
light particles take a part of the excitation energy and heavy
residua of the evaporation have  smaller energy -- due to the
momentum conservation in each consecutive two-body decay of the
de-exciting nucleus. Thus, the intermediate mass fragments emitted
from a multifragmentation could have on average larger energies than
the evaporation residua with the same mass. Due to this fact, the
evaporation should give rise to qualitatively different spectra than
emission of the same heavy particles in the multifragmentation. This
effect can explain the appearance of two qualitatively different
components of the spectra of intermediate mass fragments. Thus, the
present study concentrates on the investigation of the discussed
above hypothesis concerning properties of the spectra of
intermediate mass fragments.

Since exact calculations of multifragmentation of heavy nuclei are
practically not possible due to an enormous number of involved
partitions, and since the intranuclear cascade models, which neglect
the shell structure of nuclei might not work well for very light
nuclei (cf., \emph{e.g.}, Z. Fraenkel \emph{et al.} \cite{FRA82A}),
the choice of the target nucleus for investigation of validity of
the above proposed hypothesis should fulfill a compromise between
these two constraints. The aluminium target has been therefore
selected for the present study.

The paper is organized as follows: the experimental details are
discussed in the next section,  the analysis of the IMF data
performed by means of the intranuclear cascade (for the first stage
of the reaction) combined with evaporation and multifragmentation
(for the second step) is presented in the third section, and the
last section summarizes the obtained results.

\section{Experimental results}\label{sec:experiment}

The PISA experiment has been performed using the internal beam of
COSY (COoler SYnchrotron) of the Research Center in Juelich. The
apparatus and experimental procedure have been described in previous
publications \cite{BUB07A,BUD08A,BUD10A} thus in the present work
only details, characteristic for the studied reactions are
discussed.

Self-supporting  aluminium target of the 170 $\mu$g/cm$^2$ thickness
was bombarded by the internal proton beam of COSY. Three beam
energies were used: 1.2, 1.9, and 2.5 GeV.  To assure the same
experimental conditions for all beam energies  COSY operated in the
so called supercycle mode.  In this mode several cycles were
alternated for each requested beam energy, consisting of protons
injection from the cyclotron JULIC to COSY ring, their acceleration
with the beam circulating in the ring below the target, and
irradiating the target by slow movement of the beam in the upward
direction. Due to the application of the supercycle mode of the
target irradiation all conditions of the experiment except the
energy of the proton beam remained unchanged. This allowed to
minimize the effect of systematic uncertainties on the energy
dependence of the measured cross sections.

Double differential cross sections $d^{2}\sigma/d\Omega dE$ were
measured at seven scattering angles: 15.6$^{\circ}$, 20$^{\circ}$,
35$^{\circ}$, 50$^{\circ}$, 65$^{\circ}$, 80$^{\circ}$, and
100$^{\circ}$. The mass and charge identification of detected
particles was realized by the $\Delta$ E - E method using telescopes
consisted of silicon semiconductor detectors backed (in four cases:
15.6$^{\circ}$, 20$^{\circ}$, 65$^{\circ}$, and 100$^{\circ}$)  by a
7 cm thick CsI detector with a photo-diode readout, which were used
to detect high energy light charged particles passing through the
silicon detectors.

The $d^2\sigma/d\Omega dE$ were measured for the following
ejectiles: p, d, t, $^{3,4,6}$He, $^{6,7,8,9}$Li, $^{7,9,10}$Be,
$^{10,11,12}$B and C. Here we will concentrate on the IMF data.
Discussion on the light charged particle emission will be published
separately.

In order to allow for absolute normalization of differential cross
sections we performed a "two moving sources" fit to $d^2\sigma/
d\Omega dE$ for $^{7}$Be production in our experiment and compared
the resulting total cross section with data from ref. \cite{BUB04A}.
It was found that the angular and energy dependence of $d^2\sigma/
d\Omega dE$ can be well reproduced by a simple formula representing
the isotropic emission from two sources moving forward along the
beam direction. Each source emitting particles with Maxwellian
energy distribution was characterized by its velocity $\beta$,
temperature $T$, height of the Coulomb barrier between $^7$Be and
the source remnant – described by parameter $k$, and by emission
intensity $\sigma$ (see Appendix of ref. \cite{BUB07A} for details
of the parametrization). The $\sigma$ parameter has the meaning of
the energy and angle integrated cross section attributed to a given
source. Thus, the not normalized cross section $\sigma_{a.u.}$ for
$^{7}$Be production is equal to the sum of parameters $\sigma_1 +
\sigma_2$ for both sources. Best values of the parameters were found
by fitting simultaneously the full set of the $^7$Be spectra (seven
scattering angles). Alas, the fits lead to ambiguous results because
the experimental spectra did not cover the full energy range allowed
by kinematics. Low energy particles were not registered because of
finite low-energy detection thresholds of the telescopes.
This lack of information on the low energy part of spectra could
strongly influence the value of the energy integrated cross section
since the spectra have Maxwellian shape with its maximum lying in
the neighborhood of the energy threshold. Fortunately, it turned out
that the spread of values of $\sigma$ parameters was smaller than 10
\% among the sets of parameters which provided the same, best
$\chi^2$ values obtained for various combinations of fixed Coulomb
barrier parameters $k_1$ and $k_2$. The final values of the $\sigma$
parameters for both sources were taken as the arithmetic mean of
results obtained for equivalent quality fits, \emph{i.e.}, those
which have the same, smallest $\chi^2$ value.  The error of the
normalization factor obtained in such a way was $\approx 9\%$ for
all studied energies.  This value does not include the inaccuracy of
the literature value of the production cross section
$\sigma$($^7$Be)
\cite{BUB04A} which is believed to be smaller than 10\%.

\begin{figure}
  \centering
  \includegraphics[width=0.5\textwidth]{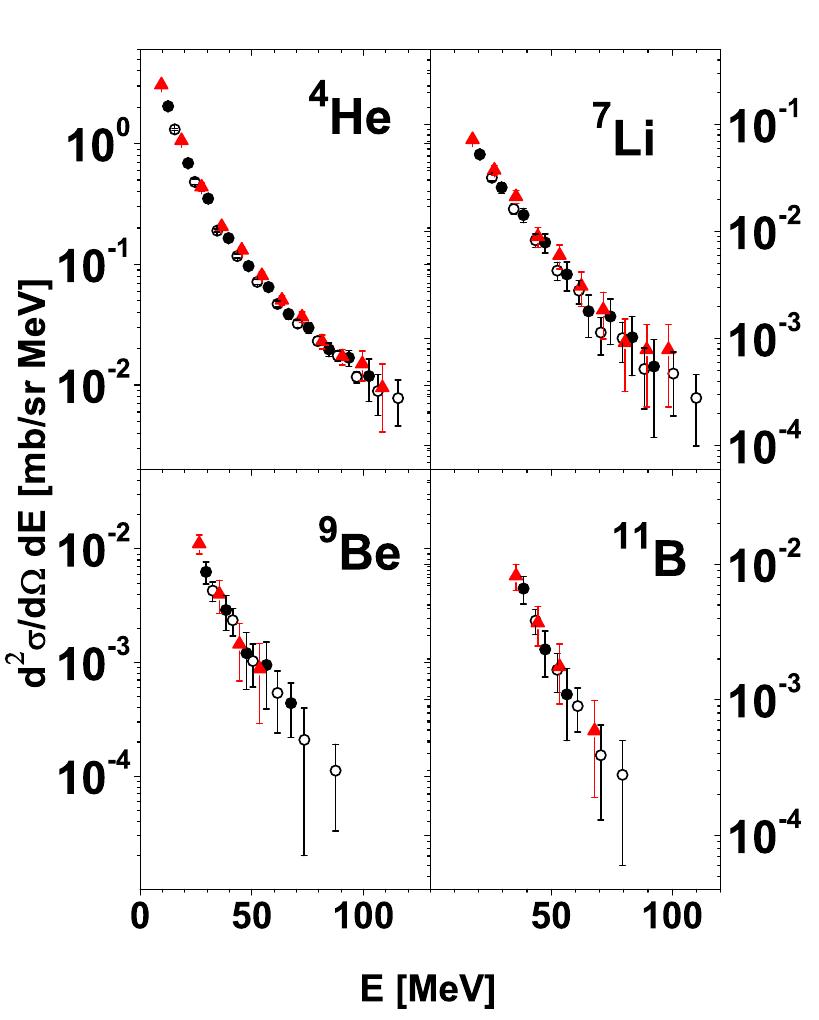}\\
  \caption{Spectra of $^{4}$He, $^{7}$Li, $^{9}$Be, $^{11}$B
   measured at 35$^{\circ}$  for p+Al collisions at three proton beam
  energies: 1.2 (circles), 1.9 (full circles), and 2.5 GeV
  (triangles). To avoid overlapping of symbols only different energy
  bins are shown for different beam energies.}\label{fig:helibebal}
\end{figure}

\begin{figure}
  \includegraphics[width=0.5\textwidth]{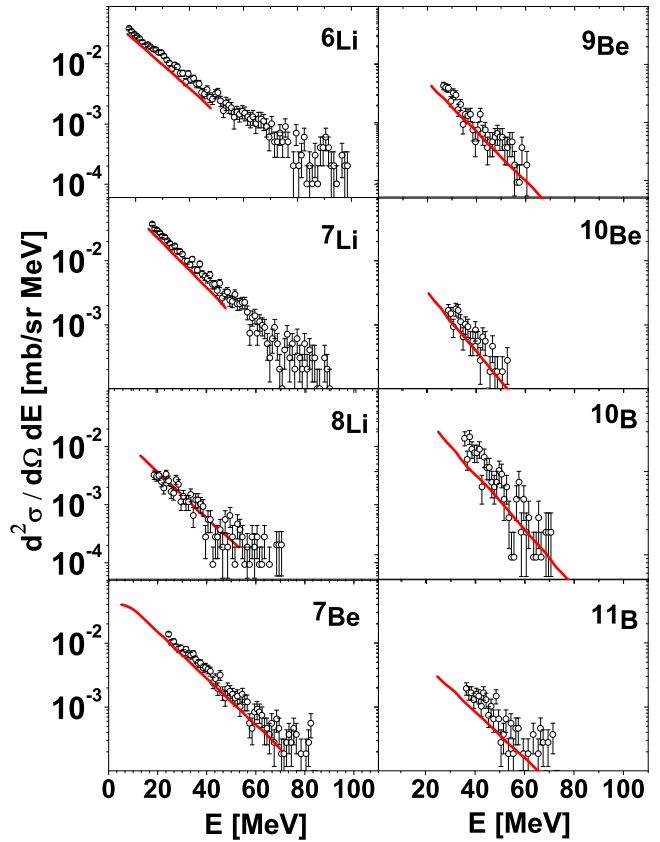}\\
  \caption{Experimental  spectra  of  Li,  Be,  and  B  from  the  present  work  (circles)  measured
at 100$^{\circ}$  in p+Al collisions at proton beam energy 2.5 GeV
and data from ref. \cite{WES78A} measured at 90$^{\circ}$  at proton
energy 4.9 GeV (lines).}\label{fig:li6b11comal}
\end{figure}

The good quality of the absolute normalization of the present
experiment is confirmed by a comparison of the obtained differential
cross sections with the data from literature.  To our knowledge only
one measurement of light charged particle spectra as well as
intermediate mass fragment spectra is present in the literature for
aluminium target at proton beam energies similar to those used in
the present study. This is the paper of Westfall \emph{et al.}
\cite{WES78A} dealing with reactions in the p+Al system investigated
at 4.9 GeV proton energy.  This energy is 2--4 times higher than
those used in the present work, however, it is known that the total
production cross sections for this target vary only slightly  at
proton beam energies larger than 1 GeV (cf. \emph{e.g.} ref.
\cite{BUB04A}). The differential cross sections measured in the
present experiment also change only slightly with the beam energy as
it is shown in Fig. \ref{fig:helibebal}. As can be seen, the cross
sections increase very slightly with the increasing beam energy.
This increase is stronger for lighter ejectiles ($^{4}$He,
$^{7}$Li), whereas the data for heavier IMFs are in the limits of
errors independent of the beam energy.

The weak energy dependence of the differential cross sections leads
to the conclusion, that it is reasonable to compare the present data
with results obtained by Westfall \emph{et al.} \cite{WES78A} at
even higher incident proton energy. Such a comparison is depicted in
Fig. \ref{fig:li6b11comal} for lithium, beryllium and boron
isotopes. As can be seen, the shapes and magnitudes of the spectra
are in excellent agreement for all ejectiles.

\section{Theoretical analysis}\label{sec:IMF}

The model calculations of the first stage of the reaction were done
by means of the INCL4.3 computer program \cite{BOU02A,BOU04A}
realizing the intranuclear cascade of nucleon-nucleon collisions.
The decay of remnant nuclei (A,Z) excited to the energy $E^{*}$ was
evaluated in the frame of two different models: the generalized
evaporation model GEM2 of S. Furihata \cite{FUR00A,FUR02A} and the
multifragmentation treated as the Fermi breakup model using the
computer program ROZPAD of A. Magiera \cite{MAG10A}.

The Fermi breakup model \cite{FER50A,GOE84A,GEANT4} assumes that the
probability $W$ of disintegration of an excited nucleus with mass
$M$ and volume $V$ into channel containing $n$ fragments with masses
$m_i$ and spins $s_i$ ($i=1,2,...,n$) is proportional to its phase
space volume \cite{GEANT4}
%
\[
W \sim \frac{g}{G}\left[ {\frac{V}{{\left( {2\pi \hbar } \right)^3
}}} \right]^{n - 1} \left( {\frac{1}{M} \prod\limits_{i = 1}^n
{m_i}} \right)^{3/2} \frac{{\left( {2\pi } \right)^{3\left( {n - 1}
\right)/2} }}{{\Gamma \left( {\frac{3(n - 1)}{2}} \right)}}
E_{kin}^{\left( {3n - 5} \right)/2}
\]
where  $g$ is the number of spin projections of $n$ fragments
$g=\prod\limits_{i = 1}^n \left( 2 s_i +1  \right)$, $G=
1/\prod\limits_{j = 1}^k n_j !$ accounts for identity of $n_j$
fragments of kind $j$ ($k$ is defined by $n=\sum\limits_{j=1}^k
{n_j}$), and $\Gamma(x)$ is the gamma function. The total kinetic
energy of $n$ fragments at the moment of breakup $E_{kin}$ is
related to the excitation energy $E^{*}$ of the decaying nucleus
(A,Z) according to the equation:
\[
E_{kin}= (E^{*} + M c^2) - \sum\limits_{i=1}^n m_i c^2 -U_n^C
\]
where $U_n^C$ represents the Coulomb interaction energy of the
fragments.

The double differential cross sections $d^2\sigma/d\Omega dE$ are
obtained in the Fermi breakup model by random generation of energies
and  flight directions of fragments over the whole accessible phase
space \cite{GEANT4}.

It should be pointed out that the Fermi breakup model
 gives  statistical method of
dealing with the multifragmentation process based on the assumption
of full equilibration of the excited system in which the energy and
linear momentum conservation is exactly taken into account. In the
following the "multifragmentation" will be understood as such a
specific model of this process.

 It was assumed that the sequential
emission of IMFs, \emph{i.e.}, the evaporation of a single particle
in each step, dominates when the remnants of the cascade are excited
to low energies whereas the remnants with higher excitation energies
split simultaneously into many pieces thus they are subject of
multifragmentation. \emph{The critical excitation energy per
nucleon} $(E^{*}/A)_{cr}$, \emph{i.e.}, the smallest value of
excitation energy per nucleon  at which multifragmentation appears
was treated in the present work as a free parameter. The only free
parameter of the Fermi breakup model - the so called
\emph{freeze-out radius}
$r_{0}$ was also fitted. 
It has the meaning of a reduced radius of the decaying, spherical
nucleus of the mass number $A$, \emph{i.e.} the volume $V$ of the
decaying nucleus is calculated as $V = 4 \pi r_0^3 A /3$.  The
default value of the $r_0$ parameter used in the GEANT4 is equal to
1.4 fm \cite{GEANT4}.

\begin{figure}
  \includegraphics[width=0.5\textwidth]{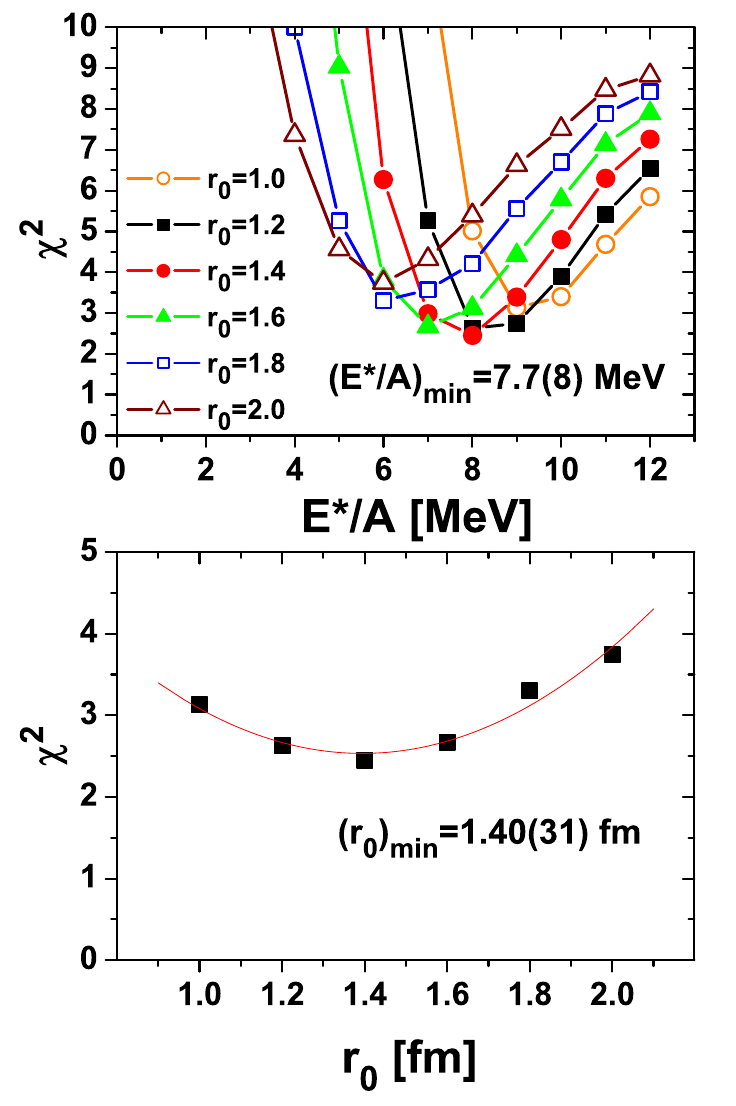}\\
  \caption{Results of the analysis for the proton beam energy equal to 1.2 GeV.
  Upper panel: chi-square values versus critical value of the excitation energy per nucleon
  $(E^{*}/A)_{cr}$ for several values of reduced freeze-out radius $r_0$ (in fm). Lower panel:
  Minimal values of the chi-square for each value of $r_0$.
  }\label{fig:ale2chi}
\end{figure}

The free parameters of the present model: freeze-out radius $r_{0}$
and critical excitation energy $(E^{*}/A)_{cr}$ were searched for by
comparison of theoretical spectra with the experimental data, for
$^{6}$He, $^{6,7,8,9}$Li, $^{7,9,10}$Be, and $^{10,11,12}$B.
Chi-square values $\chi^2$ were evaluated summing over all these
data, \emph{i.e.} over all ejectiles listed above and over seven
scattering angles for each of these ejectiles for several fixed
values of $r_{0}$ parameter and several values of critical
excitation energy $(E^{*}/A)_{cr}$. The obtained values of the
chi-square are shown in  Fig. \ref{fig:ale2chi} for proton beam
energy 1.2 GeV.

The following properties of the chi-square dependence are clearly
visible:
\begin{itemize}
\item
Broad minima of the chi-square treated as a function of the critical
excitation energy $(E^{*}/A)_{cr}$ are present for each fixed value
of the $r_{0}$ parameter. The chi-square  increases very strongly
when the critical excitation energy decreases to values smaller than
$\approx$ 5 MeV/nucleon.  This behavior points to the fact that the
nuclei at such low excitation energies are not subject of the
multifragmentation but rather  sequentially evaporate particles.

\begin{figure}
  \includegraphics[width=0.5\textwidth]{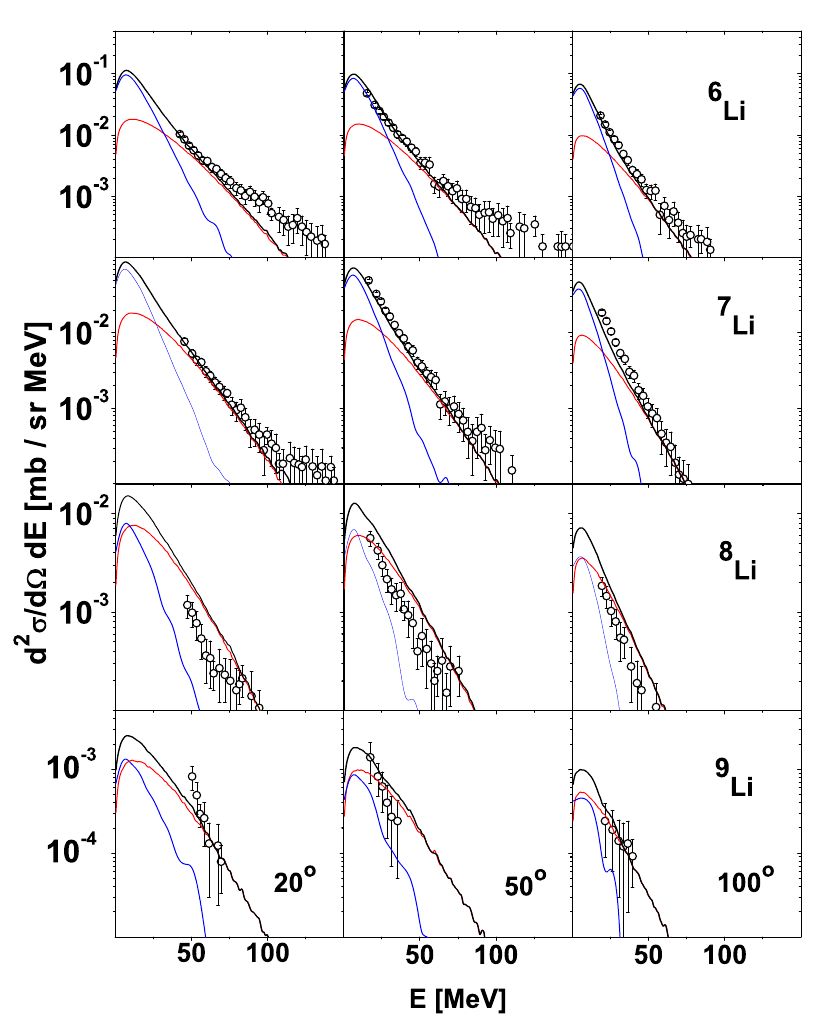}\\
  \caption{Experimental spectra (circles) measured at three scattering angles:
   20$^{\circ}$, 50$^{\circ}$, and 100$^{\circ}$ (left, middle, and right columns) for
   $^6$Li, $^7$Li, $^8$Li, and $^9$Li (upper, middle, and lower rows) for p+Al collisions
   at proton beam energy of 1.2 GeV.
   The lines represent theoretical calculations: blue (the steepest), red (the least steep), and
   black lines correspond to evaporation, multifragmentation, and
   their sum, respectively.
    }\label{fig:al2li}
\end{figure}

  \item Minimal value of the
chi-square for each value of the $r_{0}$ parameter, found from the
chi-square dependence on $(E^{*}/A)_{cr}$, is presented as a
function of $r_{0}$  in the lower panel of the Fig.
\ref{fig:ale2chi} . As can be seen, this dependence may be well
approximated by a concave parabolic function, which minimum allows
to choose the best fit value of the  $r_{0}$ parameter.
\end{itemize}

\begin{figure}
  \includegraphics[width=0.5\textwidth]{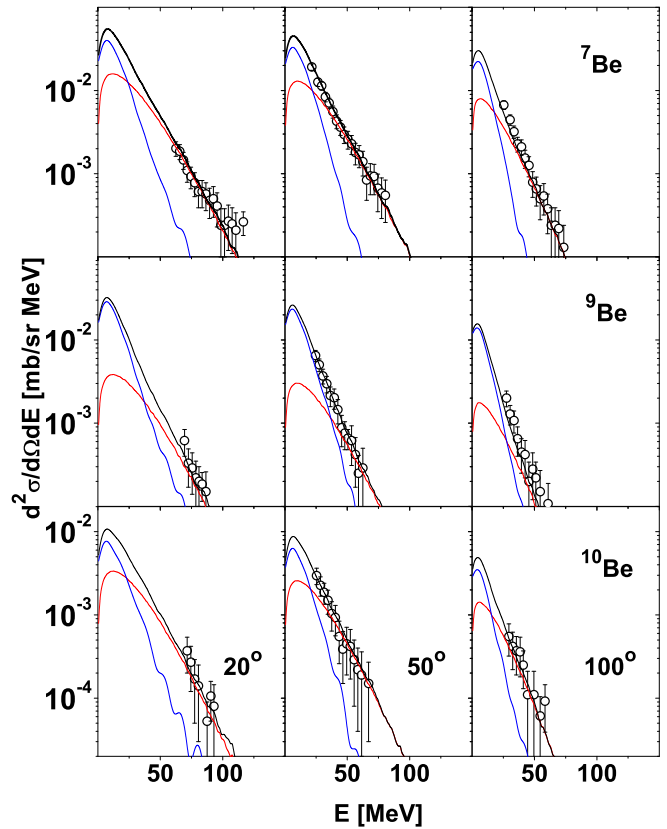}\\
  \caption{Same as in Fig. \ref{fig:al2li} but for $^7$Be, $^9$Be, and
  $^{10}$Be,
  respectively}\label{fig:al2be}
\end{figure}

The fit procedure described above has been applied to the data
measured at all three proton beam energies: 1.2, 1.9, and 2.5 GeV,
leading to the same (in the limits of errors) best fit values for
the reduced radius  $r_0$: 1.40(31) fm, 1.40(25) fm, and 1.39(43) fm
for beam energy 1.2, 1.9, and 2.5 GeV, respectively. It is worth
pointing out that the obtained values of $r_0$ parameter are equal
in the limits of errors to $r_0=$1.4 fm  assumed as the default
value in the literature \cite{GEANT4}. Thus fixing the parameter
$r_0$ at
 $r_0 = 1.40(20) \;\rm{fm}$
allows to find the best fit value of the second parameter - the
critical excitation energy $(E^{*}/A)_{cr}$ - independently for each
beam energy.

\begin{figure}
  \includegraphics[width=0.5\textwidth]{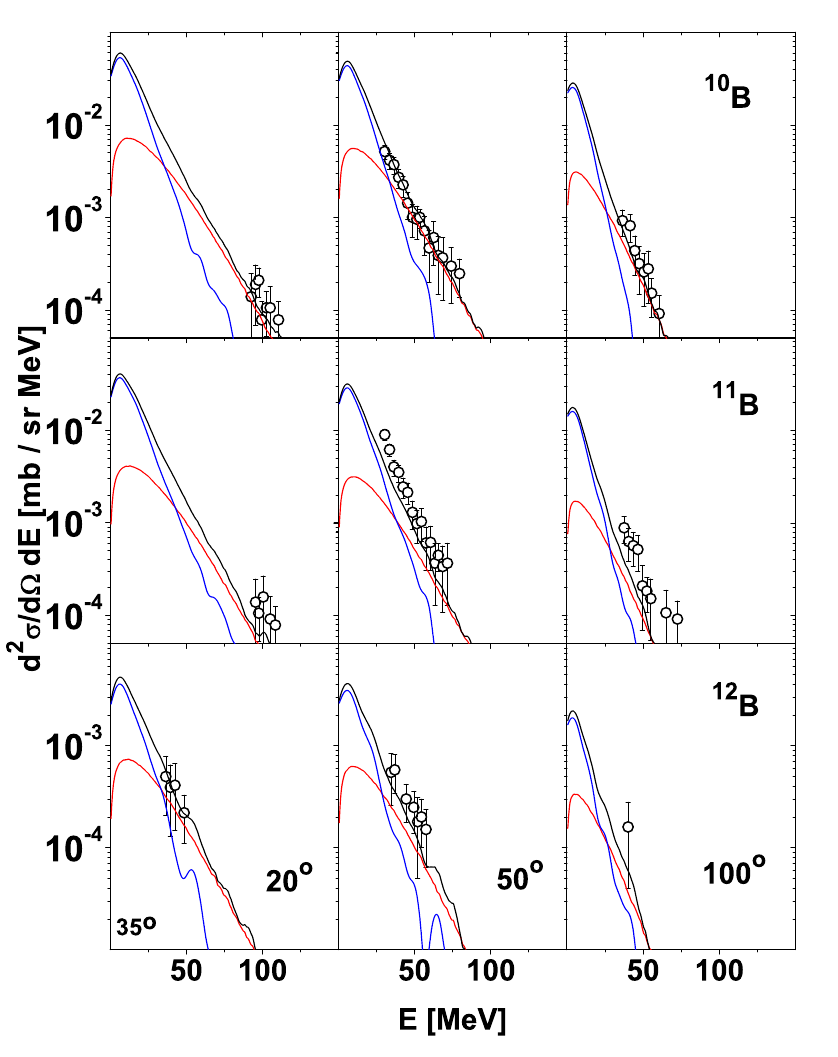}\\
  \caption{Same as in Fig. \ref{fig:al2li} but for $^{10}$B, $^{11}$B, and
  $^{12}$B,
  respectively}\label{fig:al2b}
\end{figure}

It was found that the critical excitation energy is also almost the
same in the studied beam energy range. It is equal to 7.7(8)
MeV/nucleon, 7.0(8) MeV/ nucleon, and {6.1(9) MeV/nucleon for 1.2,
1.9, and 2.5 GeV beam energy, respectively.

The quality of the description of experimental data for p+Al
collisions may be judged from inspection of Figs \ref{fig:al2li} -
\ref{fig:al2b}, where a representative sample of the data measured
at proton beam energy equal to 1.2 GeV is shown. As can be seen, the
contribution of multifragmentation to the spectra (shown by the
least-steep, red line in the figures)  is crucial for reproducing
high energy parts of the spectra of IMFs. On the other hand the
evaporation which leads to very steep shape of the spectra is not
able to describe their high energy tails but dominates at low
energies of the ejectiles. It turned out that the sum of evaporation
and multifragmentation describes very well the full spectra of
$^{6,7,8,9}$Li, $^{7,9,10}$Be, and $^{10,11,12}${B}. Note that the
fluctuations of the theoretical curves visible in the figures have
no meaning - they are only due to a limited statistics of the Monte
Carlo calculations.

 It should be emphasized that all the spectra shown here, obtained
 at 1.2 GeV proton beam energy, are reproduced
assuming the same values of the parameters: $r_0 = 1.4 \; \rm{fm}$
and $(E^{*}/A)_{cr}=7.7 \; \rm{MeV}$.

An equally good description of IMF data was achieved for higher beam
energies (1.9 and 2.5 GeV) using the same value of $r_0$ parameter
and appropriate critical excitation energies. Since the shape of the
spectra almost does not change with the beam energy, the Figs
\ref{fig:al2li} - \ref{fig:al2b} are representative for the quality
of data reproduction obtained at all studied energies.

\begin{table}
\centering \caption{\label{tab:AL_IMF_XSECS}Total production cross
sections of intermediate mass fragments for p+Al collisions at three
proton beam energies: 1.2, 1.9, and 2.5 GeV.  In the first column
the symbol of the ejectile is listed, in the second column the beam
energy is specified, and in the following columns the production
cross section due to the sequential evaporation, production cross
section due to the multifragmentation, the sum of both cross
sections, and the relative contribution of the multifragmentation to
the total cross section are presented.} \vspace*{0.5cm}
\begin{tabular}{|c|c|c|c|c||c|}
\hline
 particle & energy  & $\sigma_{GEM}$ & $\sigma_{FBM}$ & $\sigma_{Tot}$ & $\sigma_{FBM}$\\
          & [GeV]   &   [mb]         &    [mb]        &   [mb]         &    [\%]  \\
\hline
          & 1.2          & 0.57                & 1.03                 & 1.60  & 64\\
$^6$He    & 1.9          & 0.45                & 1.68                 & 2.13  & 79\\
          & 2.5          & 0.34                & 2.33                 & 2.67  & 87\\
\hline
          & 1.2          & 9.81                & 4.28                 & 14.09 & 30\\
$^6$Li    & 1.9          & 7.35                & 7.02                 & 14.37 & 49\\
          & 2.5          & 5.60                & 9.32                 & 14.92 & 62\\
\hline
          & 1.2          & 3.57                & 4.04                 & 7.61  & 53\\
$^7$Li    & 1.9          & 2.72                & 6.61                 & 9.33  & 71\\
          & 2.5          & 1.95                & 9.40                 & 11.35 & 83\\
\hline
          & 1.2          & 0.37                & 1.55                 & 1.92  & 81\\
$^8$Li    & 1.9          & 0.28                & 2.55                 & 2.83  & 90\\
          & 2.5          & 0.21                & 3.71                 & 3.92  & 95\\
\hline
          & 1.2          & 0.056               & 0.24                 & 0.296 & 81\\
$^9$Li    & 1.9          & 0.046               & 0.42                 & 0.466 & 90\\
          & 2.5          & 0.033               & 0.62                 & 0.653 & 95\\
\hline
          & 1.2          & 2.19                & 3.56                 & 5.75  & 62\\
$^7$Be    & 1.9          & 1.66                & 6.17                 & 7.83  & 79\\
          & 2.5          & 1.24                & 7.98                 & 9.22  & 87\\
\hline
          & 1.2          & 1.63                & 0.78                 & 2.41  & 32\\
$^9$Be    & 1.9          & 1.30                & 1.37                 & 2.67  & 51\\
          & 2.5          & 0.99                & 2.20                 & 3.19  & 69\\
\hline
          & 1.2          & 0.78                & 0.64                 & 1.42  & 45\\
$^{10}$Be & 1.9          & 0.63                & 1.13                 & 1.76  & 64\\
          & 2.5          & 0.45                & 1.91                 & 2.36  & 81\\
\hline
          & 1.2          & 5.86                & 1.39                 & 7.25  & 19\\
$^{10}$B  & 1.9          & 4.81                & 2.65                 & 7.46  & 36\\
          & 2.5          & 3.73                & 4.23                 & 7.96  & 53\\
\hline
          & 1.2          & 4.36                & 0.77                 & 5.13  & 15\\
$^{11}$B  & 1.9          & 3.52                & 1.50                 & 5.02  & 30\\
          & 2.5          & 2.57                & 2.71                 & 5.28  & 51\\
\hline
          & 1.2          & 0.44                & 0.15                 & 0.59  & 25\\
$^{12}$B  & 1.9          & 0.38                & 0.31                 & 0.69  & 45\\
          & 2.5          & 0.29                & 0.63                 & 0.92  & 68\\
\hline
\end{tabular}
\end{table}

The angle and energy integrated double differential cross sections
$d^2 \sigma/d\Omega dE$ obtained in the analysis described above are
collected in  Table \ref{tab:AL_IMF_XSECS}. The following
conclusions can be derived:
\begin{itemize}
  \item Evaporation cross sections decrease quickly
        with beam energy for all IMFs, on the contrary to
        increasing of
        multifragmentation cross sections.  These monotonic dependencies lead to the fast
        increase of the relative contribution of multifragmentation to
        the total cross section.
  \item The multifragmentation contribution dominates for all
        ejectiles at the highest beam energy.  It exhausts at least
        51\% of the total cross section  (for
        $^{11}$B) but for a few ejectiles it reaches even 95\% (for $^{8}$Li and
        $^{9}$Li).
  \item The sum of both contributions increases with the beam energy
        but not as quickly as the fragmentation
        cross sections themselves because of decreasing of the evaporation
        contribution.
        The ratio of the total production
        cross section at the beam energy of 2.5 GeV to that at 1.2 GeV
        varies from about 1.1 for $^{6}$Li and $^{11}$B
        to about 2.2 for $^{9}$Li.
\end{itemize}

\begin{figure}
  \includegraphics[width=0.5\textwidth]{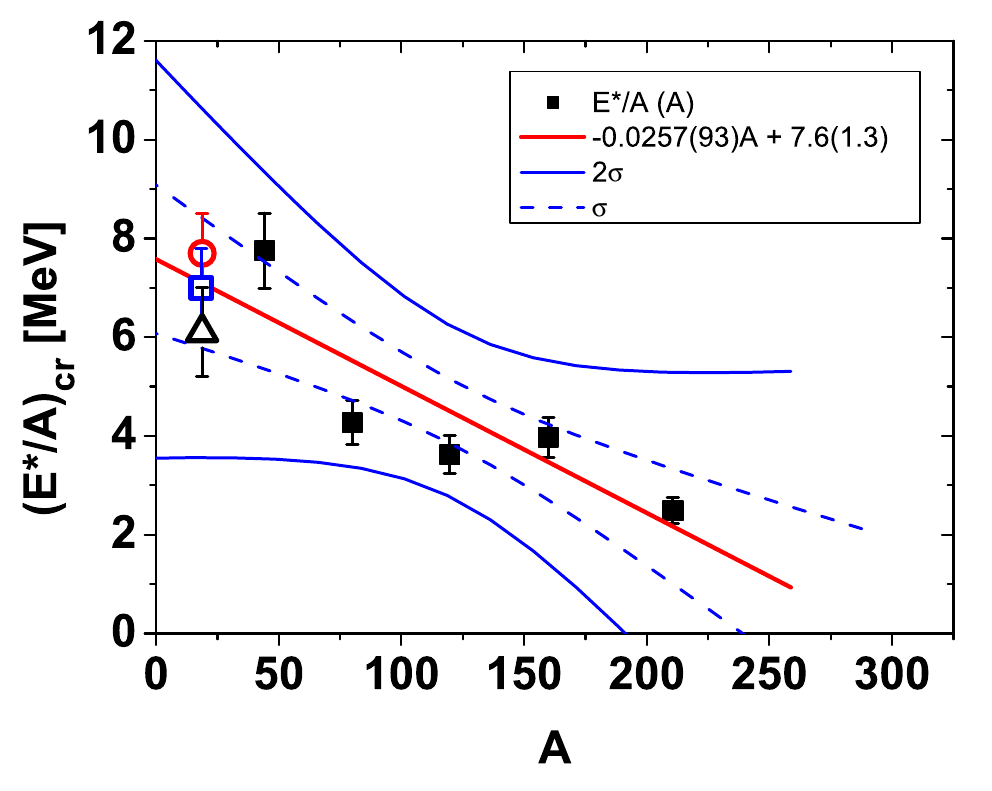}\\
  \caption{The full squares represent the compilation of critical
  excitation energies per nucleon as a function of mass number of decaying nuclei
  obtained from the study of caloric curves \cite{NAT02A},
  the solid straight line shows the linear regression of these
  data whereas the dashed and solid hyperbolas correspond to one- and two- standard deviation
  confidence intervals of the regression line, respectively.  The empty symbols  depict
  the critical excitation energy values used in the present analysis
  for p+Al collisions; circle, square, and triangle correspond to 1.2, 1.9, and 2.5 GeV
  beam energy,  respectively.}\label{fig:natalall}
\end{figure}

It was checked whether the parameter of the present model,
\emph{i.e.} the critical excitation energy per nucleon
$(E^{*}/A)_{cr}$ is compatible with the critical excitation energy
found from the analysis of the caloric curves.  For this purpose the
values of the critical energy obtained in the present investigation
were compared with those from a compilation of caloric curve results
published by Natovitz \emph{et al.} \cite{NAT02A}. The target mass
dependence of the critical excitation energy obtained from the
analysis of the caloric curves by Natovitz \emph{et al.} is
presented in Fig. \ref{fig:natalall} together with critical energies
found from p+Al collisions investigated in the present work. The
former results were averaged in Ref. \ref{fig:natalall} over thirty
units of mass of decaying nuclei and were attributed to the central
value of each mass region.  The present results correspond to
multifragmentation of an ensemble of excited nuclei with the average
mass $<A> = 18.7$ and with the standard deviation of masses
$\sigma(A)= 2.6$.
It is clear that the present values of the critical excitation
energy fit perfectly to the compilation of data obtained from the
study of caloric curves. The values of the critical energies from
the present study lay for all three proton beam energies inside the
confidence interval of one standard deviation around the regression
line approximating the mass dependence of the critical energy per
nucleon.  
%

\section{Summary}\label{sec:summary}

In the present work the hypothesis was investigated, claiming that
the energy and angular distributions of the double differential
cross sections $d^2\sigma/d\Omega dE$ for intermediate mass
fragments from inclusive measurements of reactions proceeding in the
p+Al system at GeV proton beam energies can be reproduced by a
reaction model in which a specific two-step mechanism is involved.
In the first stage of the reaction a cascade of nucleon-nucleon and
pion-nucleon collisions appears leaving an ensemble of the excited
residual nuclei. These nuclei decay in the next stage of the
reaction according to two competing scenarios: (i) The nuclei with
the excitation energy per nucleon smaller than some critical value
$(E^{*}/A)_{cr}$ - treated as a parameter, decay by a sequential
emission of evaporated particles, (ii) the nuclei excited to
energies above the limiting value $(E^{*}/A)_{cr}$ are subjects to
multifragmentation. The first of these mechanisms is responsible for
populating the low energy part of the spectra, whereas the second
leads to emission of also higher energy intermediate mass
fragments.\\

The excellent agreement of the calculated within such a model
differential cross sections $d^{2}\sigma/d\Omega dE$ with the
experimental data confirms the validity of the assumed
hypothesis.\\

The investigations performed in the present work provide also the
following conclusions:
\begin{itemize}
  \item
        It was found that all intermediate mass fragments are
        products
        of the second stage of the reaction with a relative contribution
        of multifragmentation of the remnants of the nucleon-nucleon
        cascade systematically increasing with the proton beam
        energy.\\

  \item
        The only free parameter of the calculations mentioned above was the
        critical excitation energy $(E^{*}/A)_{cr}$ which decides which option of the
        decay of the excited remnants of the nucleon-nucleon
        cascade should be chosen.  The second parameter which could
        in principle vary, \emph{i.e.} the reduced freeze-out radius $r_0$,  appeared
        to be independent of the beam energy and to be equal to 1.4 fm, \emph{i.e.} to the
        default value used in the Fermi breakup
        model (see \emph{e.g.} ref. \cite{GEANT4}).\\
  \item
        The value of the critical energy found from the present analysis of
        the inclusive spectra is compatible with those obtained
        from the study of the caloric curves.  Therefore, it may be concluded that the
        investigation of  intermediate mass fragment spectra
        obtained
        in  inclusive measurements may be used as an alternative
        method to studying of the caloric curves
        for the extraction of the critical excitation energy per
        nucleon.\\
\end{itemize}


Successful description of the  intermediate mass fragment emission
from p+Al collisions in the 1.2 - 2.5 GeV proton beam energy range
suggests that a similar picture of the reaction mechanism might be
also appropriate for heavier nuclear systems as, \emph{e.g.}, p+Ni
or p+Au where another model of the IMF emission has been proposed
\cite{BUB07A,BUD08A,BUD09A,BUD10A}. 
\begin{acknowledgments}

The technical support of A.Heczko, W. Migda{\l}, and N. Paul in
preparation of experimental apparatus is greatly appreciated. This
work was supported by the European Commission through European
Community-Research Infrastructure Activity under FP6 project Hadron
Physics (contract number RII3-CT-2004-506078), and HadronPhysics2
(contract number 227431). 
One of us (M.F.)
acknowledges gratefully financial support of Polish Ministry of
Science and Higher Education (Grant No N N202 174735, contract
number 1747/B/H03/2008/35).

\end{acknowledgments}

\end{document}